\pdfoutput=1
\documentclass[osajnl,twocolumn,showpacs,superscriptaddress,11pt]{revtex4-1} 
\usepackage{amsmath,amssymb,graphicx}
\usepackage{multirow}
\usepackage{setspace}
\usepackage{caption}
\usepackage{subcaption}
\usepackage{hyperref}

\begin{document}

\title{Optical characterization of gaps in directly bonded Si compound optics using infrared spectroscopy}

\author{Michael Gully-Santiago}\email{gully@astro.as.utexas.edu}
\author{Daniel T. Jaffe}
\affiliation{Department of Astronomy, The University of Texas at Austin, Austin, TX, 78712, USA}

\author{Victor White}
\affiliation{NASA Jet Propulsion Laboratory, Pasadena, CA, 91109, USA}

\begin{abstract}
Silicon direct bonding offers flexibility in the design and development of Si optics by allowing manufacturers to combine subcomponents with a potentially lossless and mechanically stable interface. The bonding process presents challenges in meeting the requirements for optical performance because air gaps at the Si interface cause large Fresnel reflections. Even small (35 nm) gaps reduce transmission through a direct bonded Si compound optic by 4\% at $\lambda = 1.25 \; \mu$m at normal incidence.  We describe a bond inspection method that makes use of precision slit spectroscopy to detect and measure gaps as small as 14 nm.  Our method compares low finesse Fabry-P\'{e}rot models to high precision measurements of transmission as a function of wavelength.  We demonstrate the validity of the approach by measuring bond gaps of known depths produced by microlithography.
\end{abstract}


\maketitle 

\section{Introduction}

Crystalline silicon is an excellent material for infrared optics. Si has a high refractive index ($3.55-3.45$ from $\lambda = 1150-2500\;$nm at room temperature, \cite{2006SPIE.6273E..77F}) and superb transmission from slightly longer than the equivalent wavelength of its band gap ($\sim$1.15 $\mu$m) to about 6.5 $\mu$m \cite{PhysRev.108.268, PhysRev.78.178} and in the far-IR \cite{doi:10.1117/12.323764}.  It also has adequate transmission  ($\alpha \sim 1$ cm$^{-1}$) for many purposes in the 18$-$30 $\mu$m region \cite{doi:10.1117/12.323764}.  There is an extensive industrial infrastructure for the production of ultrapure Si and for polishing and lithographic patterning and etching.

As with a number of glasses and crystalline optical materials, two planar or conformal silicon surfaces can form strong physical bonds.  One advantage of this direct Si$-$Si bonding of optical parts is that it facilitates the manufacture of complex optical subsystems, for example double-convex aspheres, combinations of lenses, prisms, and transmission gratings, and pairs of gratings oriented at right angles \cite{2012SPIE.8450E..2TV, 2010SPIE.7739E.123G}.  A sufficiently close bond is optically lossless, an important consideration not only for its effect on throughput but also because close lossless bonds can eliminate concerns about optical ghosts. Other advantages of bonding stem from the difficulties encountered in antireflection coating a high index material in the near and mid-infrared:  There is a limited choice of index-matching materials, in particular at longer wavelengths.  Systems often are used at cryogenic temperatures and problems with differential thermal expansion, brittleness, and hygroscopic properties can restrict the choice of coatings even further. It is hard to get good antireflection performance across broad spectral ranges and at a range of incidence angles.

There are several good reviews of Si bonding that recount the history of the field and describe process alternatives, metrology, and problems \cite{1998AnRMS..28..215G,Masteika2014}.  The foundation paper on Si bonding came in 1986 \cite{1986JAP....60.2987S}. After that, there was a sequence of papers that used IR imaging to look for evidence of defects in the bond. Stengl et al. (1988) and G{\"o}esele et al. (1995) \cite{1988JaJAP..27L2364S, 1995ApPhL..67.3614G} directly monitored the bond propagation with time-lapse IR imaging. Lehman et al. 1989 and G{\"o}esele et al. \cite{1989JaJAP..28L2141L, 1995ApPhL..67.3614G} demonstrated gap-free bonding under a vacuum in a micro- cleanroom, in which bonding occurs immediately after a rinse-spin-vacuum cycle.  Other authors investigated the growth of bubbles at bond interfaces with time and the elimination of some or all of these bubbles after annealing bonded parts at temperatures of $800-1100\;^\circ$C \cite{Horn2009, Masteika2014}.

The chemistry of the bonding interface began to be understood with infrared absorption spectroscopy and multiple internal reflection spectroscopy\cite{feijoo1994}, which lent credibility to the idea of distinct temperature phases and evidence for Si$-$H bonds\cite{1995ApPhA..61..101R}.  A robust picture of the chemistry as a function of annealing temperature emerged \cite{1996JaJAP..35.2102R, 1998AnRMS..28..215G}.  Other variables potentially affecting bond strength were studied, for example surface roughness\cite{JJAP.37.4197}, surface topography \cite{2001JOptA...3...85G}, surface preparation\cite{1996ApPhL..68.2222T}, annealing time \cite{2000JAP....88.4404H}, ambient pressure and substrate thickness \cite{1995ApPhL..67..863G, 2007ApOpt..46.6793H}, and moisture and defects \cite{2001JAP....89.6013L}.  Notably, \cite{JJAP.37.4197} found that surface roughness above 1.3 nm RMS results in poor bonding quality in their sample of wafers etched with Argon.  Greco \emph{et al.} examined the role of surface topography in determining the bonding strength of thick optical surfaces.  These authors showed that van der Waals interactions by themselves were consistent with their available experimental evidence, but they could not rule out other forces and phenomena.  The dearth of measurements of the mean separation between contacted surfaces hindered the unambiguous identification of the forces at play in bonded, rough surfaces \cite{2001JOptA...3...85G}.

If Si$-$Si bonds are to be a useful part of the optical designer's toolbox, the bonds must have transmission losses similar to or lower than the concatenated losses of two antireflection coated surfaces over the operating wavelength (typically $< 4$\%).  For Si$-$air gaps, the full Fresnel loss is $\sim30\%$ per surface, much larger than the $4-7\;\%$ per surface of optical materials with refractive indices $1.5-1.7$. A 4\% transmission loss specification for Si translates to physical gaps or bubbles with an axial extent of $ < 35\;$nm.  If the optical layout does not eliminate pupil ghosts from the flat Si surfaces, there are additional requirements on the reflectivity of individual bubbles in the bond. While completely bonded Si parts will be completely lossless at the interface, problems on different scales with different physical origins can lead to less-than-perfect bonds.  On the smallest scales, imperfect cleaning can leave small voids where particulates hold the bonding planes apart \cite{Mitani1990} or where hydrocarbons can serve as catalysts for gap formation.  Even in the best of cleanrooms, particles at the optically significant $10-20\;$nm scale can be present in relatively high abundance. Figure 5 of Cooper 1986 \cite{doi:10.1080/02786828608959094} shows the surface flux (particles cm$^2$ s$^{-1}$) vs. particle size ($\mu$m). There will be roughly $10^4$ times more 10 nm sized particles than 1 $\mu$m sized particles.  Another possible source of gaps at the bond is inherent roughness in the initial surfaces that leads to a failure to conform \cite{2001JOptA...3...85G}.  On larger scales, gaseous hydrogen, which is a byproduct of the hydrophilic bonding process, can migrate into the bond and form local bubbles \cite{Masteika2014}.  

Most of the bonding literature concerns itself with wafer bonding, and not thicker optical substrates.  In bonding wafers to each other or to thicker optically polished Si disks, one or both of the partners can conform as the bonding front propagates.  In optics, however, both of the bonding partners will be thicker and more rigid.  When bonding two thick substrates, small differences in the shapes of the polished surfaces could lead to topological inconsistencies that prohibit some areas from bonding.  If enough pressure is placed on the substrates and if the bond strength is high enough relative to the stiffness of one or the other component, however, the surfaces will conform and bond.  Subsequent annealing will strengthen the bond.

Verification that an optical bond meets its requirements and diagnosis of possible problems in the bonding technique require accurate measurement of the lateral and axial dimensions of any defects.  Since silicon is opaque in the optical range, the most straightforward way to detect defects in the bond is to take infrared images.  Fresnel losses will cause unbonded regions to appear darker than regions where the bond is perfect.  Newton rings appear in the image where the axial extent of the gap is a half-wavelength or more.  X-ray topography \cite{Mitani1990} exploits the change in optical density along lines of sight to allow precise high spatial resolution measurements of the spatial extent of gaps with axial dimensions as small as a few nanometers but not of the axial extent itself. In ultrasound microscopy, the density change at the gap interface produces reflections of sound waves and enables mapping the lateral gap dimensions but also produces only limited information about the axial dimension \cite{2000RScI...71.1869G}.

We describe here a new technique for measuring the axial extent of small gaps, founded upon the ability to measure transmission as a function of wavelength with high accuracy.   A recent generation of stock near-infrared spectrophotometers like the Cary5000 from Agilent offer $0.2\;\%$ precision and good repeatability.  We use the wavelength dependence of the transmission and a model of the gap as a closely-spaced Fabry-P\'{e}rot etalon to measure axial gaps down to a few tens of nanometers in size.  In order to verify the applicability and accuracy of the technique, we used microlithography to create small artificial gaps of known size in wafers that we then bond and measure.

For all types of gaps and bubbles, our technique needs to provide an accurate assessment of the transmission loss at the operating wavelength.  For finite-sized bubbles, measurement of the axial dimension helps us to assess the efficacy of high-temperature annealing \cite{Horn2009, Masteika2014} in the removal of hydrogen bubbles.  A wavelength-dependent transmission variation can also indicate the presence of large numbers of gaps that are too small to be resolved spatially, as one might expect if there are many very small contaminants present along the interface.  Our technique must be able to characterize gaps with losses of a few percent.  A $2\%$ loss at $1.25\;\mu$m corresponds to a gap of $>25\;$nm at normal incidence and the gap size for a loss of this magnitude scales linearly with wavelength.  We therefore need to be able to measure transmission as a function of wavelength to a fraction of a percent, including all systematic errors, for devices operating near $1\;\mu$m and to a somewhat looser standard for devices operating at longer wavelengths.

\section{Theory: Modeling the Si$-$Si interface gap as a low finesse Fabry-P\'{e}rot etalon}
\label{secTheory}

Our modeling technique exploits the large Fresnel reflection \cite{2001opt4.book.....H} at the Si$-$vacuum interface.  We model a small Si$-$Si interface gap between two almost-bonded parts as a low-Finesse Fabry-P\'{e}rot etalon \cite{2007fuph.book.....S}.  The key property of an etalon is that it has a cavity enclosed between two smooth and parallel reflective surfaces where the requirements for smoothness and parallelism depend on the reflective finesse.  For the narrow air gaps we consider here and for the relatively modest reflectivity of a Si$-$vacuum interface, most gaps at Si$-$Si bonds will form effective etalons over most of their area.  The transmission through a Fabry-P\'{e}rot depends on the wavelength of light, the reflectivity of the etalon sidewalls, and the size of the gap.  The reflectivity is given by Fresnel's law at normal incidence:

\begin{eqnarray}
R = \frac{(n-1)^2}{(n+1)^2} \label{Eq:FresnelR}\\
F \equiv \frac{4R}{(1-R)^2} \label{Eq:coeffF}
\end{eqnarray}

\begin{figure}[!htbp]
\includegraphics[width=0.95\columnwidth]{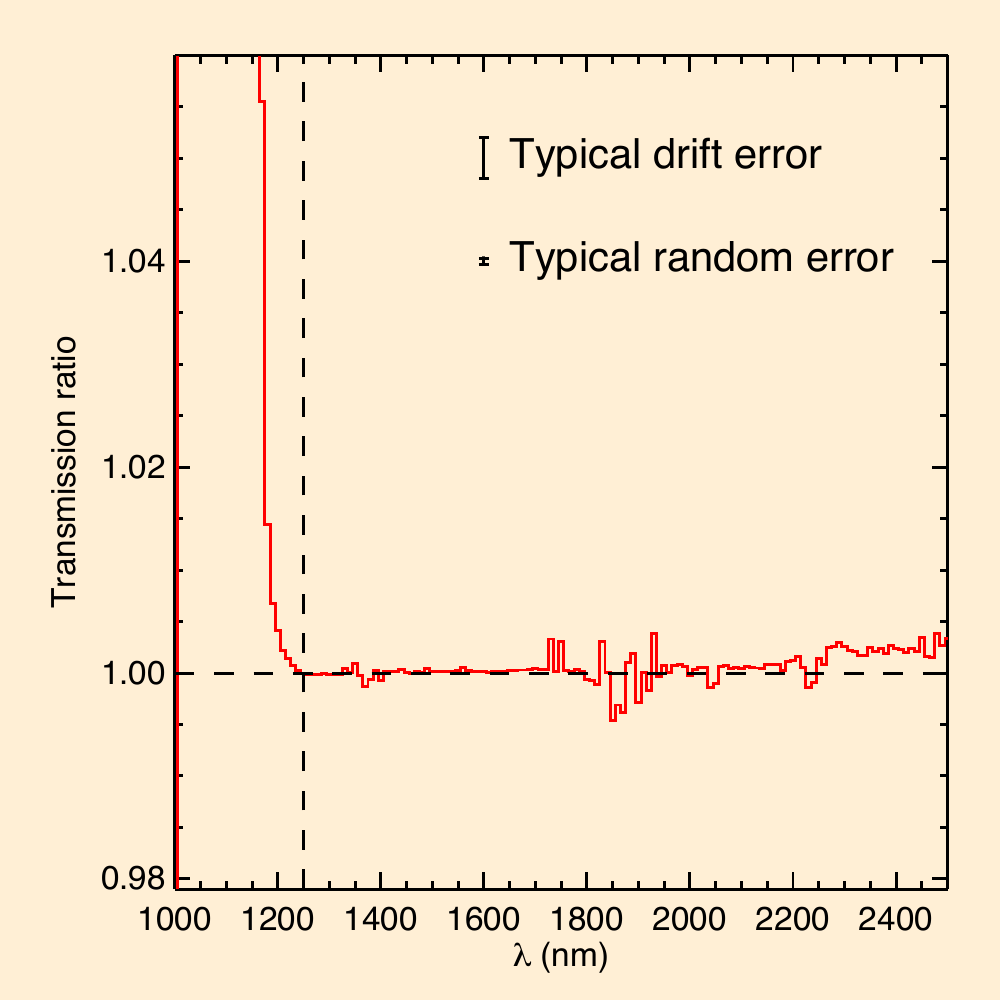}
\caption{Ratio of transmission of a 0.5 mm thick Si wafer to that of a 3.0 mm thick Si puck.  \label{figSiAbsorbfig} We detect Si absorption shortward of $\lambda$ = 1250 nm.  The individual measurements have correlated uncertainties attributable to lamp drift at the level of 0.2\%.  The random uncertainties are typically about 0.02\% per sample, though some wavelength regions (e.g. 1700-1900 nm) demonstrate much greater per sample uncertainties.  Absorption longward of 1250 nm (vertical dashed line) is negligible.}
\end{figure}

We define the coefficient of finesse\cite{2007fuph.book.....S}, $F$ (Equation \ref{Eq:coeffF}), in the customary way to encapsulate the Fabry-P\'{e}rot etalon's dependence on reflectivity.  $F$ has a value of 2.5$-$2.6, depending on wavelength.  We assume room temperature for all calculations of refractive index.  In this work, we assume transmission is at normal incidence and the refractive index of the gap is 1.0.  Si absorbs negligibly longward of $\lambda$ = 1250 nm, which we verified experimentally by comparing the transmission of Si reference samples of different thicknesses (Figure \ref{figSiAbsorbfig}).

\begin{figure}[!htbp]
\includegraphics[width=0.85\columnwidth]{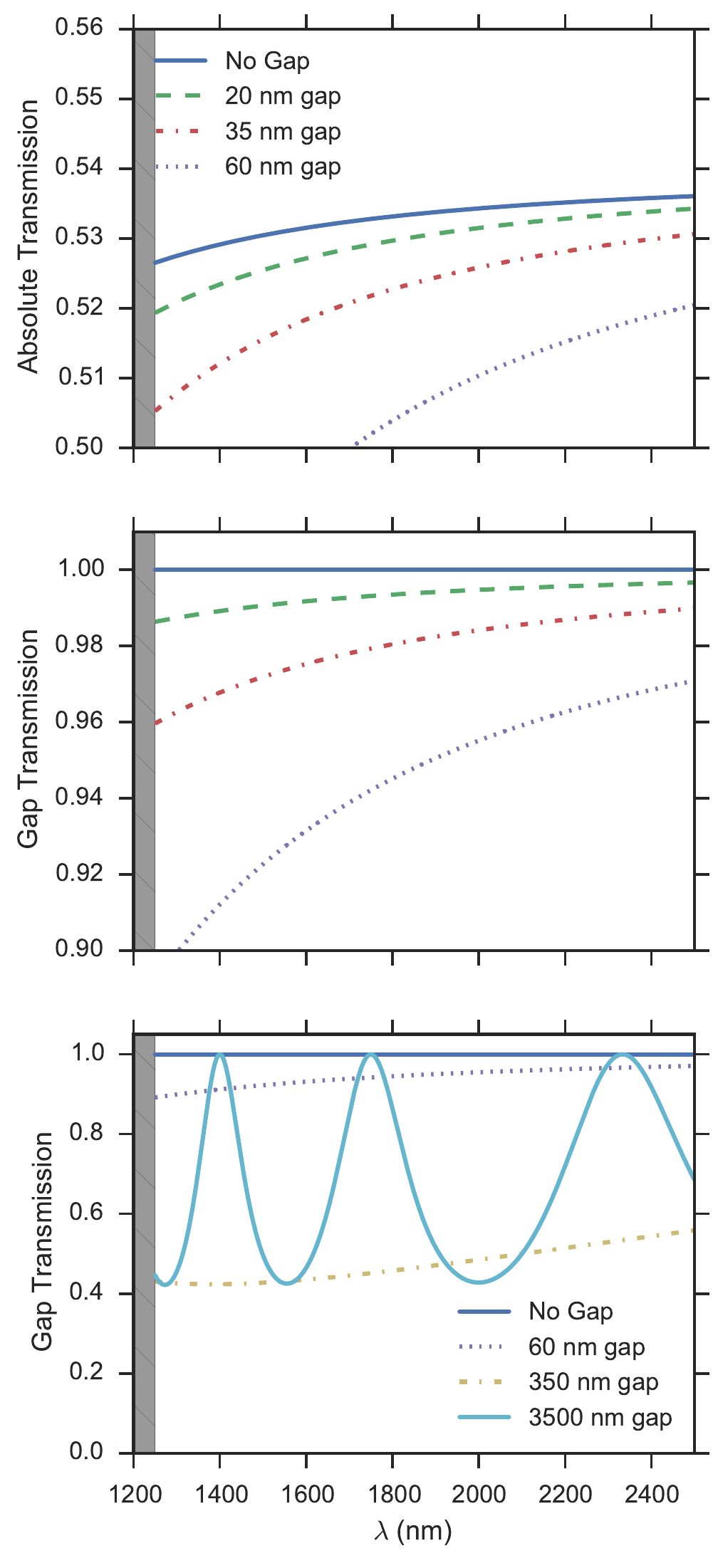}
\caption{\label{figAbsoluteTrans}The family of normal-incidence curves for transmission as a function of vacuum wavelength for various gap sizes in bonded Si optics.  \emph{Top panel}: Absolute transmission through a Si substrate with no gap (top denim-blue solid curve), and small gaps of axial extent 20 nm (green dashed curve), 35 nm (red dash-dotted curve), and 60 nm (purple dotted curve).  The absolute transmission through Si depends on wavelength since the refractive index depends minutely on wavelength.  \emph{Middle panel}: The ``Gap Transmission'', $T_g$. The gap sizes and colors are the same as in the top panel.  \emph{Bottom panel}: ``Gap Transmission'' for larger gaps, spanning 0$-$3500 nm.  The 3500 nm sized gap (cyan solid line), demonstrates Fabry-P\'{e}rot etalon fringes.}
\end{figure}

A pair of bonded silicon wafers or pucks with a small gap forms three coupled cavities with the air gap as the central cavity.  The axial extent of the silicon cavities are large enough---and typical measurement spectral resolution low enough---that the Si cavities reflect and transmit light incoherently at their entrance and exit interfaces.  In the air gap, however, the gap size is small enough that the cavity it forms can always be treated as a coherent structure.  We adapted coherent wave transfer techniques \cite{2007fuph.book.....S} for use with incoherent interactions \cite{2002ApOpt..41.3978K} to deal with the cavities within thick Si substrates.  The details appear in Appendix \ref{sec:Append-IMRTMM}.  We computed analytic equations for total transmission (including multiple internal reflections) for two scenarios: first a double side polished (DSP) Si sample with no gap, and second a pair of bonded Si samples with a gap thickness $d$.  The DSP Si sample with no air gap has a transmission equal to:

\begin{eqnarray}
T_{DSP} = \frac{2n}{1+n^2} \label{eqnAbsDSPtrans}
\end{eqnarray}

which has an average value of about 53\%.  The top line in the upper panel of Figure \ref{figAbsoluteTrans} plots the value of $T_{DSP}$ from 1200 to 2500 nm.

We derive the absolute transmission of bonded Si substrates with a gap in Equation \ref{eqn:Tetalon} in the Appendix.  We divide the absolute transmission by $T_{DSP}$ to isolate the effect of the gap.  We call this \emph{DSP-normalized} transmission the gap transmission $T_{g}$.

\begin{eqnarray}
T_{g} = \frac{n^2+1}{2 n F \sin ^2(2\pi \frac{d}{\lambda})+n^2+1} \label{eqFP}
\end{eqnarray}

In the limit $\lim_{d \to 0} T_g \rightarrow \frac{n^2 + 1}{n^2 + 1} = 1 $ and the gap approaches 100\% transmission.  Figure \ref{figAbsoluteTrans} shows a plot of Equation \ref{eqFP} for gap sizes, $d$, of 0, 20, 35, 60, 350, and 3500 nm.  The bottom two panels show the gap transmission based on Equation \ref{eqFP}, the value of the throughput with the losses from the outer Si cavities divided out.  For the very largest gap, large fringes appear as the cavity resonance of the air gap modulates the transmission.  Below a $\lambda/2$ gap size, the gap manifests itself as a wavelength-dependent diminution of the transmission.  For the smallest ($35-60$ nm) gaps, there is a $2-6\%$ decrease in gap transmission from $2400\;$nm down to $1250\;$nm.  The magnitude and wavelength-dependence of this transmission spectrum provide information about the axial extent of the gap.

We also define a mixture model in which the observed normalized gap spectrum $T_{obs}$ is composed of a sum of a spectrum through a gap of axial extent $d$, and a spectrum through a perfect bond:

\begin{equation}
	T_{mix} = f\;T_{g} + (1-f) \label{eqnMix}
\end{equation}

where $f$ is the areal fill factor of the region exhibiting a finite gap size, and $1-f$ is therefore the areal fill factor of the perfectly bonded area.  $T_g$ is defined in Equation \ref{eqFP}.  The only free parameters in Equation \ref{eqnMix} are $d$ (included in $T_g$) and $f$.

\section{Embedded gaps of known axial and lateral extent}

We evaluated our metrology method experimentally by embedding gaps of known sizes in the interface between two bonded Si substrates and measuring the effects of these gaps on IR transmission.  The substrate thicknesses ranged from 0.5 to 3.3 mm.  Both $d$ and $f$ are fixed in our direct bonded Si wafers with synthetic gaps, and are listed in Table \ref{tbl_meshPatterns}.  Synthetic embedded gaps therefore offer an excellent test of our ability to recover the gap axial extent $d$ and areal fill factor $f$ (Section \ref{secResults}).

The three important characteristics of gaps are their axial extent, their lateral areal extent, and the areal fill factor of ensembles of such gaps.  To fabricate our test samples, we used photoresist lithography to pattern Si substrates, and we bored holes with inductively coupled plasma etching.  Tables \ref{tbl_experiments} and \ref{tbl_meshPatterns} give information about the hole patterns and depths, as measured with Veeco NT9100 Optical Profiler for the small depths, and Dektak stylus profilometry for the large depths.

\begin{figure*}[!htbp]
\includegraphics[width=0.75\textwidth]{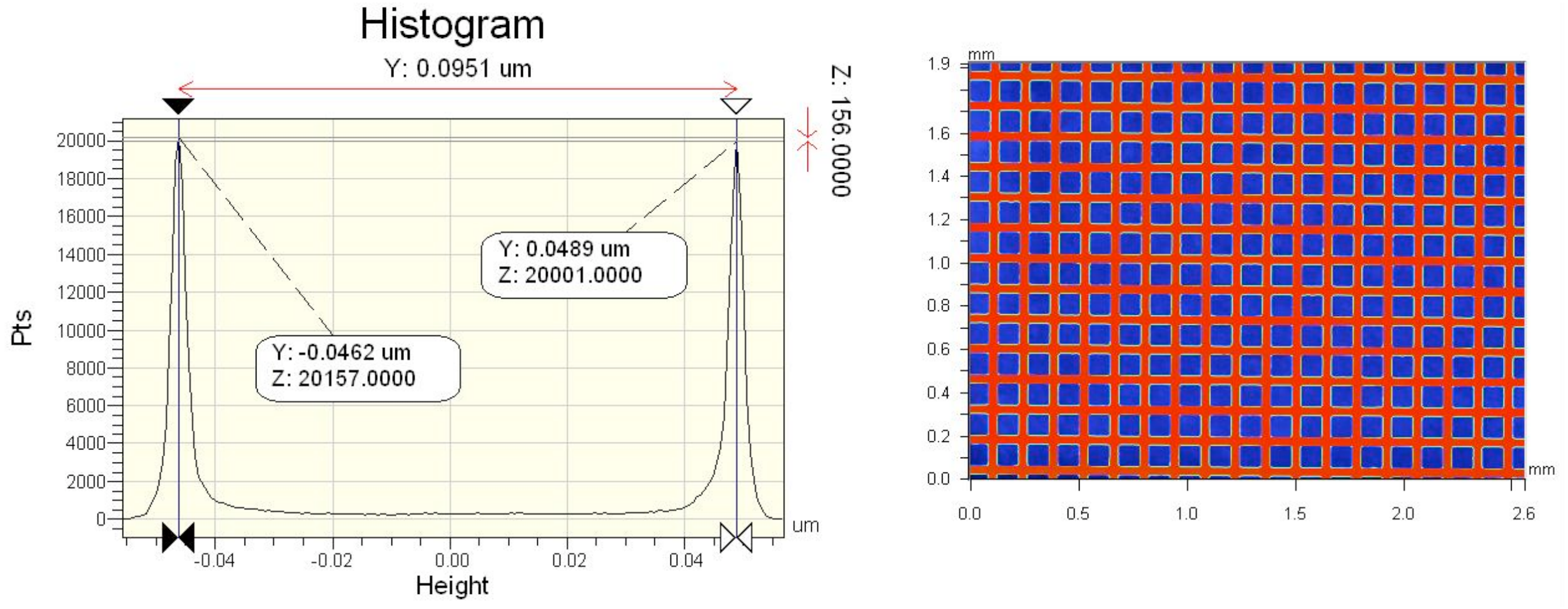}
\caption{
\label{figVS20pattern}
Veeco optical profiler surface plot and histogram for part VS20-21 in the fine pattern section.  The 100 mm diameter VS20-21 is comprised of two thick silicon wafers.  We patterned three meshes with coarse, medium, and fine boxes, each with 50\% fill factors.  The depth was $95 \pm 5\;$nm as indicated by the well-separated peaks in the Veeco profilometry histogram.  The fill factor for this and all mesh patterns was 50\%.  The purpose for these patterns was to have gaps of known axial extent against which we could test the optical metrology technique we describe in this article.}
\end{figure*}

The fill factor is the pattern area covered by gaps relative to the total pattern area.  We achieved gap sizes of $49-4000\;$nm on four substrates with three meshes with coarse, medium, and fine boxes, each with 50\% fill factors.  We did not verify the delivered fill factor, but since high precision lithography is very reliable, we assign no uncertainty to the difference between the delivered fill factor and the designed fill factor.  For measurements taken with the Veeco NT9100 Optical Profiler, uncertainties were constructed by inspecting the histogram of topology, as shown in Figure \ref{figVS20pattern}.  The stylus profilometry measured values were assigned an uncertainty of 5\%, based on our experience with other etched parts.  Table \ref{tbl_experiments} lists descriptions for our bonded substrates.

\begin{table*}[!htbp]
\caption{UTexas Si bonding experiments \label{tbl_experiments}}
\begin{center}
    \begin{tabular}{ c c c c c}
    \hline
    Name & Diameter & Thickness & Pattern & Pattern Depth\\ 
    -  & mm & mm & - & nm \\
        \hline
    VG02   & 100 & 3.3 $\times$ 2 &  None  & - \\
    VG03   & 100 & 3.3 $\times$ 2 &  Hole \& Petal & 4000$\pm$200 \\
    VG09-12   & 75   & 1 $\times$ 2 & Mesh C & 49 $\pm$6 \\
    VS20-21   & 100 & 0.8 $\times$ 2 &  Mesh C, M, F & 95 $\pm$5 \\
    \hline
    \end{tabular}
\end{center}
\end{table*}

\begin{table*}[!htbp]
\caption{Patterned gap properties \label{tbl_meshPatterns}}
\begin{center}
    \begin{tabular}{ c c c c }
    \hline
    Pattern & Fill Factor & feature size & bulk description \\ 
    - & \% & $\mu$m & - \\ 
    \hline
    Hole   & 100     &  -         & 25 mm diameter hole \\     
    Petal  & $\sim5$ & $\sim$2000 & 2 mm wide lines \\         
    Mesh F & 50      &         40 & 100 $\mu$m square holes, Fig. \ref{figVS20pattern}\\ 
    Mesh M & 50      & 200        & 500 $\mu$m square holes\\ 
    Mesh C & 50      & 620        & 1500 $\mu$m square holes\\     
        \hline
    \end{tabular}
\end{center}
\end{table*}

To minimize the interfacial particle density, we cleaned the surfaces before bonding with standard cleaning procedures of solvents in a megasonic.  The surface roughness was typically about 2 nm, as measured with a Veeco NT9100 Optical Profiler.  We also measured the large-scale surface flatness of the pucks with an interferometer.  We found a typical peak to valley surface flatness of 2 waves ($\lambda=632.8\;$nm) over the central 50 mm diameter.  After cleaning, we applied MHz frequency oxygen plasma ashing.  We soaked the wafers in DI water then dried with N$_2$.  The patterned Si substrate is bonded to a blank (unpatterned) Si substrate.  We pressed the Si substrates together from the center to the outside.

\begin{figure*}[!htbp]
\includegraphics[width=0.5\textwidth]{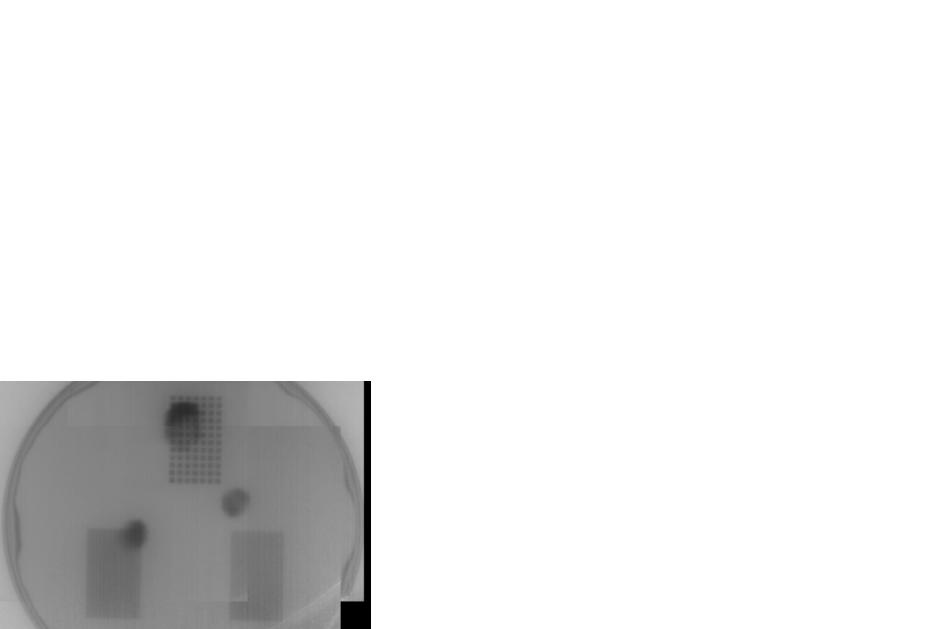}
\caption{
\label{figVS2021_IR_image}
IR image of part VS20-21. Three rectangular mesh areas are apparent (Table \ref{tbl_meshPatterns}).  The coarsest mesh, C, shows substructure, while the medium and fine meshes are indistinguishable at the low resolution of the image.  The three dark spots are bubbles at the interface of the direct bonded Si samples.}
\end{figure*}

We first looked for large gaps in our bonded substrate pairs detectable as ``IR bubbles'' \cite{1992JEMat..21..669M}, with IR imaging.  The combination of the absorption of Si, the lamp spectrum, and the detector responsivity led to an effective bandpass from approximately $1.15-1.6\;\mu$m.  The field of view was smaller than the 100 mm diameter wafers, so we dithered the sample.  Figure \ref{figVS2021_IR_image} shows an IR image of bonded sample VS20-21.  For this image, we flat-fielded the detector by imaging and dividing out a homogeneously illuminated white screen. Despite the flat-fielding, detector non-uniformities and vignetting are perceptible in our image as vertical stripes.  Mosaic-stitching errors are also perceptible.  The IR bubbles were conspicuous in the images despite detector artifacts.

We also used IR imaging to inspect VG09-12 and VG15-13.  We detected IR bubbles in all bonded Si samples.  The observable bubble areal density varied from about 4 per 100 mm diameter bonded wafer pair to over 20 per bonded wafer pair.  The individual IR bubble areas were as small as a few square millimeters up to $400\;\mathrm{mm}^2$.  The structure within some of the images of IR bubbles showed Newton's rings, revealing the presence of gaps larger than about $\lambda/2$ up to $\sim 15 \lambda$.  As expected, the unbonded DSP wafers show no such IR bubbles.  No attempt at quantitative measurement was made on the IR images due to evidence for a non-linear detector response.  When IR imaging was available, it was used to target IR spectroscopy toward locations displaying the presence or absence of IR bubbles, as desired.  In other cases when IR imaging was not available, spectra were taken at random positions on the wafer.

\section{Data}

We measured transmission spectra through the bonded Si samples with an Agilent Cary 5000 UV-Vis-NIR spectrophotometer.  Table \ref{tabCary5000pars} lists the typical measurement settings and tool performance.

\begin{table}[!htbp]
\caption{Summary of Cary 5000 measurement parameters \label{tabCary5000pars}}
\begin{center}
\begin{tabular}{ c c }
\hline
        Parameter (Units) & Value \\ 
\hline
        Spectral sampling interval (nm) & 2.0,10.0 \\
        Spectral resolution (nm) & 5 \\
        Typical measurement range (nm) & 1200-2500 \\
		Mean level drift uncertainty (\%) & 0.20 \\
		Random uncertainty (\%) & 0.02 \\
 		Beam size (mm $\times$ mm) & $2 \times 8$ \\
    \hline
    \end{tabular}
\end{center}
\end{table}

\subsection{Comparing the Fabry-P\'{e}rot model to data}

At this stage, \emph{any} regression technique can be employed to compare the Fabry-P\'{e}rot model to data.  Least squares regression and other popular methods can be applied easily.  

Our measurement uncertainty is dominated by \emph{bias}, not \emph{variance} \cite{2013sdmm.book.....I}.  The bias is similar to a mean-level drift, which may be attributable to monochromator lamp luminosity drift.  The variance is the random sample-to-sample measurement residual remaining once the bias term is removed.  In our experiments, the bias term has an amplitude of 0.2\%, whereas the variance has an amplitude of 0.02\%.

Since least squares regression assumes zero bias, we instead elected to apply a Gaussian process regression technique that can handle biased data \cite{2013sdmm.book.....I,DFMgp,rasmussen2006gaussian}.  Given the relatively small bias amplitude (0.2\%), least squares regression and Gaussian process regression might yield similar solutions for the gap size, but least squares regression would drastically underestimate the \emph{uncertainty} associated with the derived value of gap size\cite{DFMgp}.

\subsection{MCMC Gaussian process regression}

We constructed the likelihood function \cite{2013sdmm.book.....I}.  The Gaussian process takes into account the unknown but finite covariance structure by introducing a covariance matrix, $\boldsymbol{C}$ whose elements represent the correlations of each sample $i$ with every other sample $j$.  We parameterize the covariance matrix elements in the following way:

\begin{equation}
	C_{ij} = \sigma^2_{i}\delta_{ij}+a^2\exp{(-\frac{(\lambda_i-\lambda_j)^2}{2s^2})} \label{eqnGPkernel}
\end{equation}

where $\delta_{ij}$ is the Kronecker delta, and $\sigma_i$ are the independent measurement uncertainties on the $i^{\mathrm{th}}$ data point. The parameter $s$ controls the correlation length, and $a$ controls the amplitude of the correlated noise.  We experimented with different values for $\sigma_i$.  We estimated the $\sigma_i$ from the dispersion around mean-subtracted, detrended transmission spectra of DSP samples.  We do not know $a$ or $s$ and they are, in general, different for each measurement-- they are \emph{nuisance parameters}, and are ultimately marginalized out.  They arise from the behavior of the spectrometer, and their interpretation is immaterial.  The value of $a$ should be close to the mean drift reported in Table \ref{tabCary5000pars}.  The value of $s$ could be small (tens of nm) to capture local variation attributable to atmospheric air absorption, or large (hundreds or thousands of nm) to capture global features attributable to lamp drift.  We allow both $a$ and $s$ to be free parameters.

We set our prior probability distribution functions \cite{2013sdmm.book.....I}, $\ln{p(d,f,a,s)}$ as uniform over a wide range of values for $\ln{a}$, $\ln{s}$, $d$, and $f$, consistent with visual inspection of each spectrum.  The final un-normalized posterior probability function, expressed as a natural logarithm is:

\begin{eqnarray}
	& &\ln{p(d,f,a,s|\lambda, T_{obs}, \sigma_i)} \propto  \nonumber\\
	& &\,\,\,\ln{p(d,f,a,s)} -\frac{1}{2}\;\boldsymbol{r^\intercal}\boldsymbol{C^{-1}}\boldsymbol{r} -\frac{1}{2}\;\ln{\det{\boldsymbol{C}}} \label{eqnPosterior}
\end{eqnarray}

where the residual vector $\boldsymbol{r}$ is the data $T_{obs}$ minus the model $T_{mix}$.  We produce posterior samples with Markov Chain Monte Carlo (MCMC).  Specifically, we implemented \texttt{emcee}\footnote{\url{https://github.com/dfm/emcee}}\cite{emcee}, with 32 walkers, hundreds of burn-in samples, and 600-1000 iterations.  We initialized the walkers in the vicinity of our best guess for parameters based on visual inspection of the spectra.  Figures \ref{figVG03full}, \ref{figVG03part}, and \ref{figVG12} show examples of posterior samples including corner plots of the interesting physical parameters $d$ and $f$.

\section{Results of infrared spectroscopy of directly bonded Si}
\label{secResults}

\begin{figure}[!htbp]
        \centering
        \begin{subfigure}[b]{0.5\textwidth}
              \includegraphics[width=\textwidth]{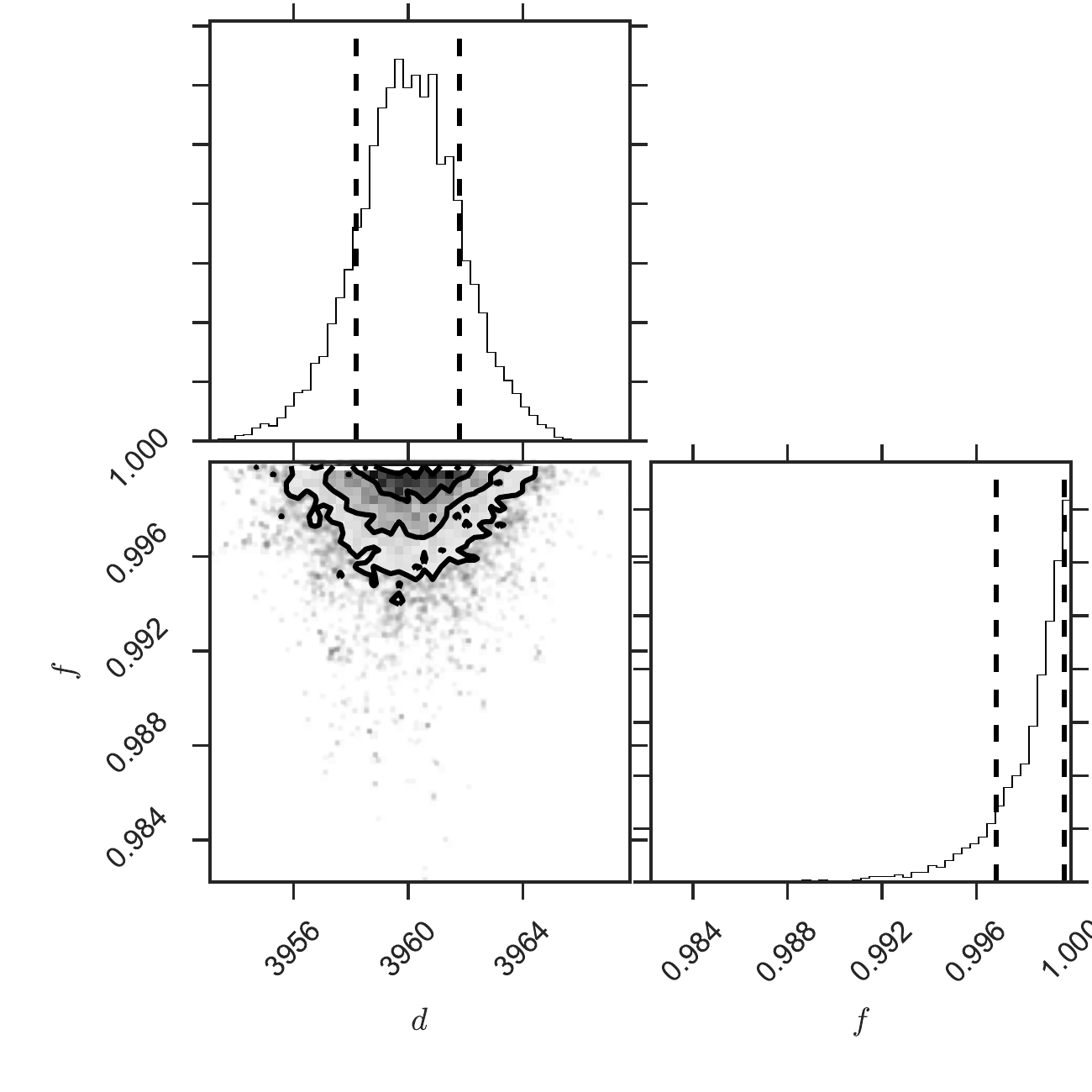}
              \caption{Fitted gap size $d$ and fill factor $f$ for VG03 hole region}
		\label{figVG03_corner}
        \end{subfigure}

        \begin{subfigure}[b]{0.5\textwidth}
                \includegraphics[width=\textwidth]{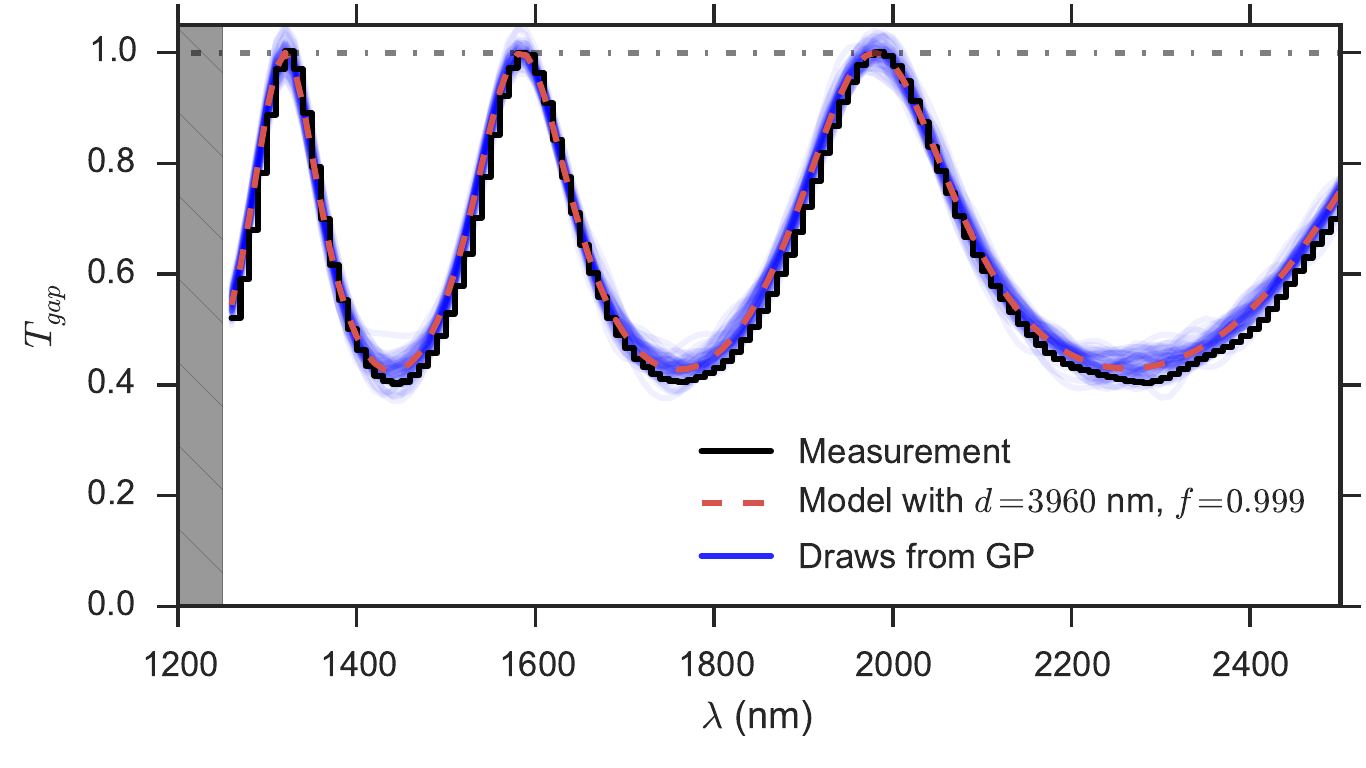}
                \caption{Measured spectrum and fit of VG03 hole region}
                \label{figVG03_f100}
        \end{subfigure}
\caption{ \emph{Top panel:} Corner plot of MCMC samples in our non-linear model fit of the transmission spectrum of Si optic part \# VG03.  The physical parameters of the model are the axial extent of the gap $d$, and the areal fill factor $f$ of the gap over the measurement region.  The MCMC model fit also includes nuisance parameters $a$ and $s$ (not shown) for the Gaussian process covariance model.  \emph{Bottom panel:}Transmission through bonded Si sample VG03, normalized by the transmission of a DSP Si wafer. See the text for details on the sample, setup, and analysis.  The measured spectrum is the black stepped line.  The red dashed line shows the best model fit with $d$ = 3960 $\pm$ 2 nm, consistent with 100\% fill factor.  The blue band shows the collective locations of 60 random draws from the Gaussian process model.  Wavelengths short-ward of 1250 nm (gray band) were not included in the fit, since Si is absorptive there.\label{figVG03full} }
\end{figure}

\begin{figure}[!htbp]
        \centering
        \begin{subfigure}[b]{0.5\textwidth}
              \includegraphics[width=\textwidth]{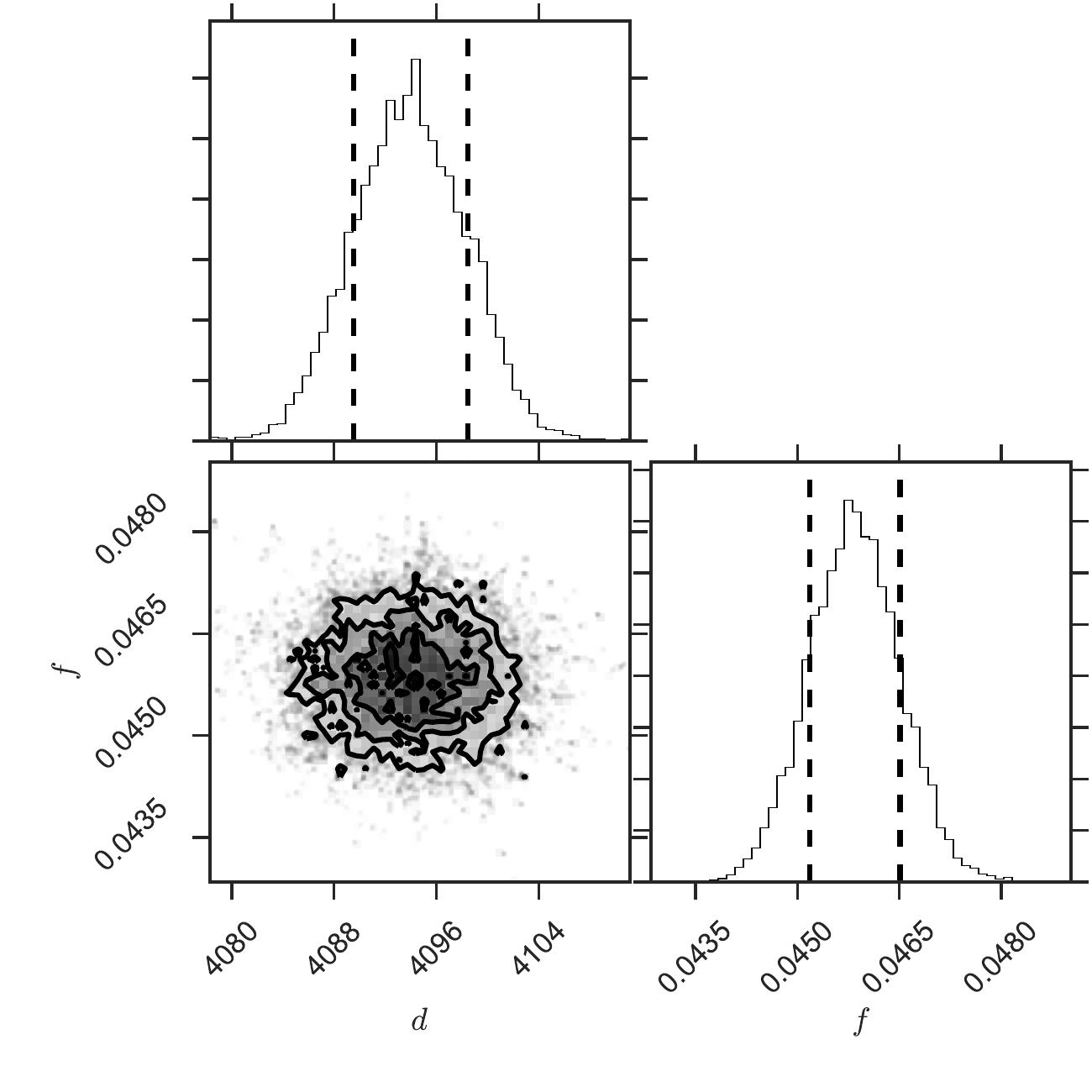}
              \caption{Fitted gap size $d$ and fill factor $f$ for VG03 petal region}
		\label{figVG03p2_corner}
        \end{subfigure}

        \begin{subfigure}[b]{0.5\textwidth}
                \includegraphics[width=\textwidth]{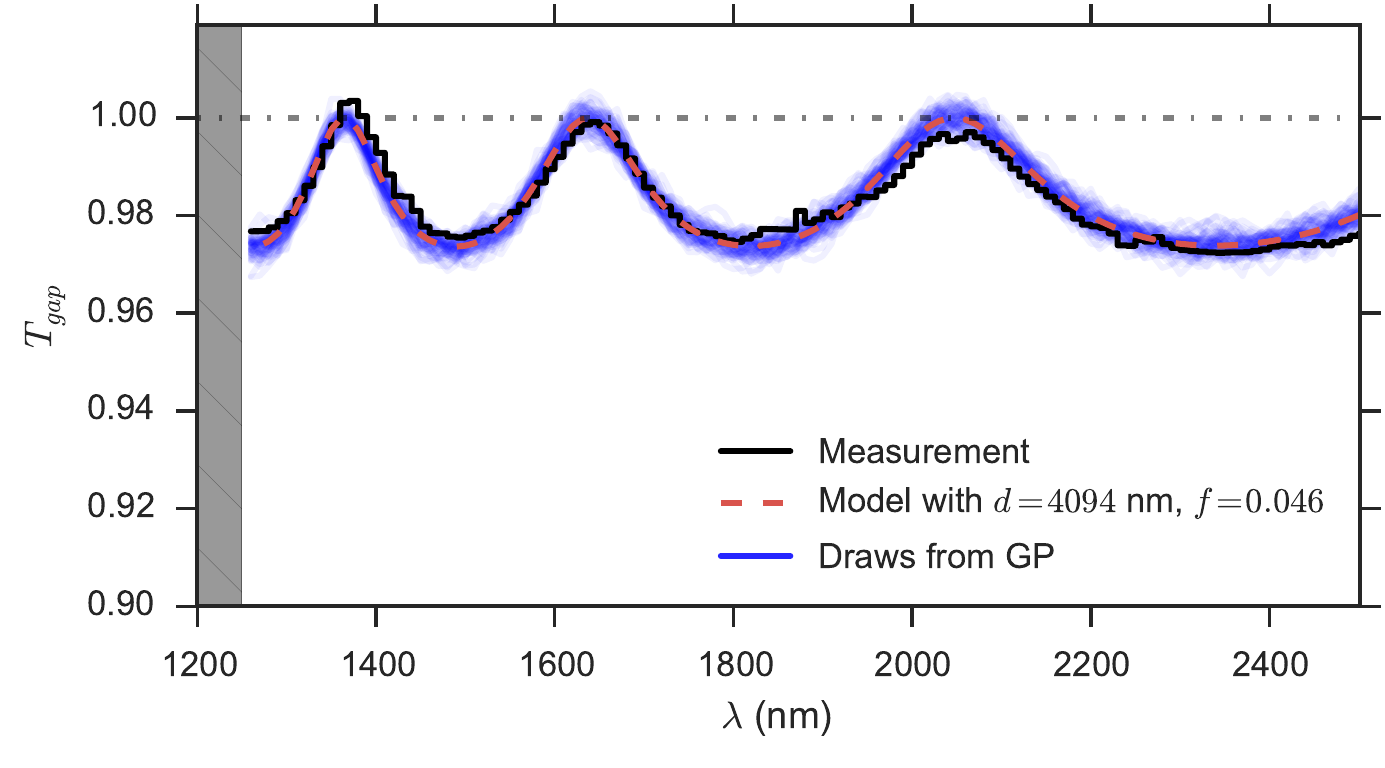}
                \caption{Measured spectrum and fit of VG03 petal region}
                \label{figVG03_f045}
        \end{subfigure}
\caption{ Same as Figure \ref{figVG03full}, but the measurement was in a region where $<5\%$ of the area exhibited the $\sim4\; \mu$m gap.  The measured spectrum in this region shows subdued fringes. The model derived axial extent is 4094 $\pm$ 5 nm covering a fill factor of 4.58 $\pm$ 0.06$\%$. \label{figVG03part}}
\end{figure}

Figure \ref{figVG03full} shows a parameter determination for the measurement of a bond over a deep ($4000 \pm 200\;$nm) 100\% fill factor gap in part VG03.  The analysis yields a best-fit consistent with 100\% fill factor, and a depth of $3960 \pm 2\;$nm, well within the uncertainty of the physical measurement (Table \ref{tbl_experiments}).  The spectrum in this region shows characteristic Fabry-P\'{e}rot interference fringes.  The only tunable parameters of the model are the gap axial extent, $d$ and fill factor $f$.  No scaling has been applied to the curves in Figure \ref{figVG03full}.  Figure \ref{figVG03part} shows a fit for a region of the same bonded pair where a nominally $4000\;$nm deep, 2 mm wide trench crosses the field (about 1 mm $\times$ 10 mm) scanned by the Cary 5000.  Analysis of this transmission curve gave a best fit depth of $4094 \pm 4\;$nm and a pattern fill factor of $4.6 \pm 0.1\;\%$, both consistent with the part geometry.  These two results show that the formalism performs as expected on resonant gaps several wavelengths deep.

\begin{table*}[!htbp]
\caption{Inferred gap sizes and fill factors \label{tbl_DerivedGapSizes}}
\begin{center}
    \begin{tabular}{ c c c c c }
    \hline
    Name & Region & Predicted $d$ & Measured $d$ & $f$ \\
    -  & - & nm & nm & - \\
    \hline
    DSP & -    &   0.0  & $3^{+8}_{-2}$ & $0.260^{+0.456}_{-0.230}$\\
    VG09-12 & Off-Mesh    &   0.0  & $11^{+21}_{-7}$ & $0.25^{+0.46}_{-0.22}$\\
    VG03 & Hole    &   4000 $\pm$ 200  & $3960^{+2}_{-2}$ &  $0.999^{+0.001}_{-0.002}$\\
    VG03 & Petal   &   4000 $\pm$ 200  & $4094^{+4}_{-4}$ &  $0.046^{+0.001}_{-0.001}$\\
    \hline
    \end{tabular}
\end{center}
\end{table*}

\subsection{Spatially resolved measurements across the face of a sample}
\begin{figure}[!htbp]
    \centering
    \begin{subfigure}[b]{0.35\textwidth}
        \includegraphics[width=\textwidth]{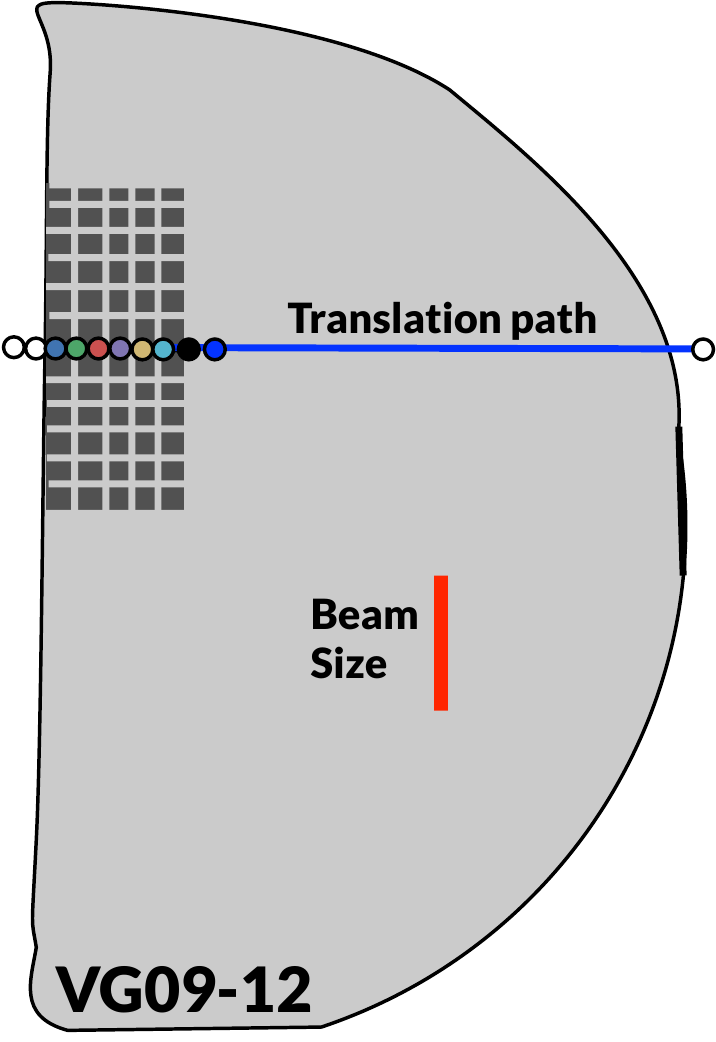}
        \caption{Illustration of VG09-12 sample transport accessory measurement positions. \label{figVG0912_STA_illus} }
    \end{subfigure}
    \begin{subfigure}[b]{0.45\textwidth}
        \includegraphics[width=\textwidth]{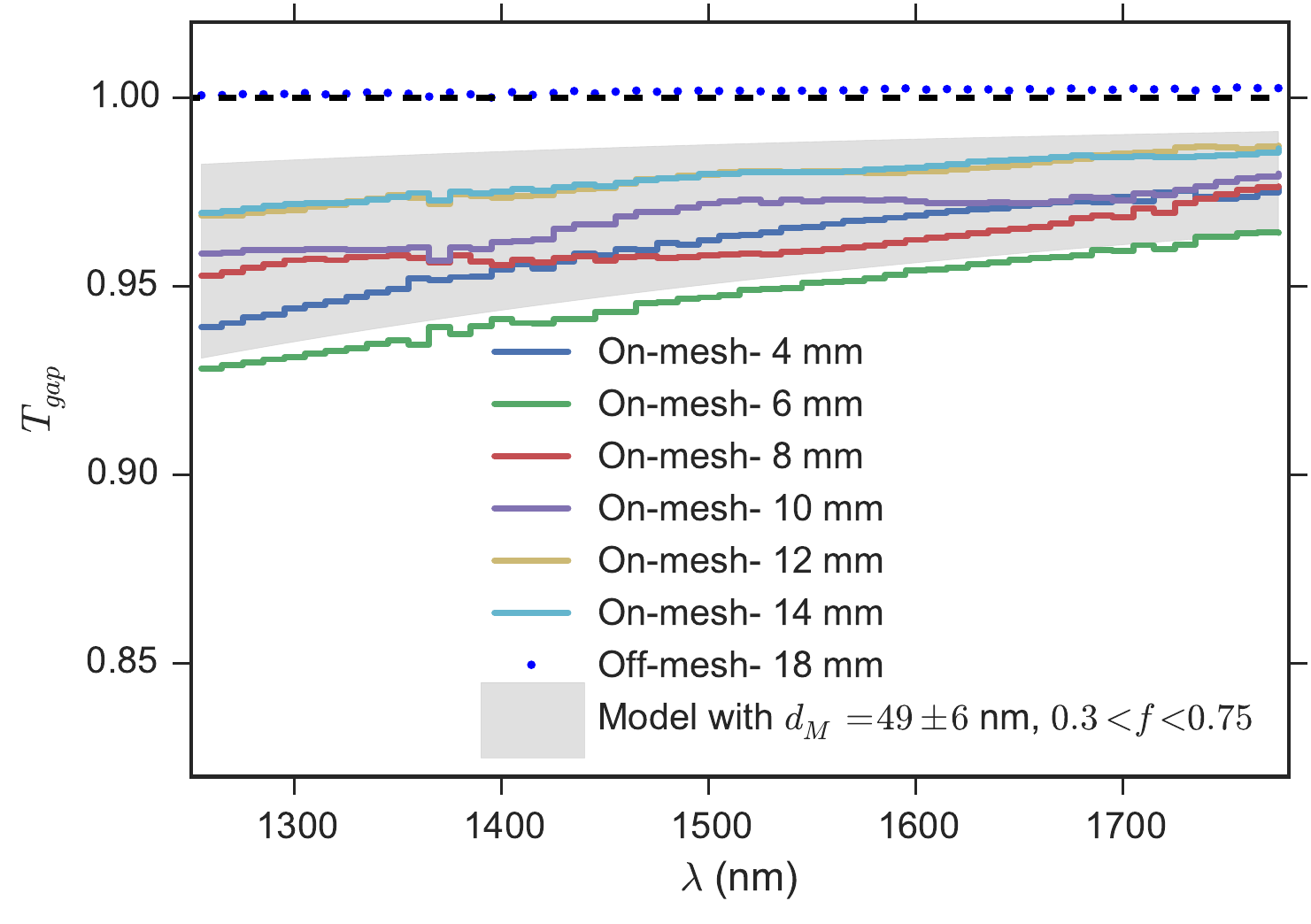}
        \caption{Measured gap transmission spectra of VG09-12 on- and off- mesh \label{figVG0912}}
    \end{subfigure}
\caption{Predictions and measurements of the coarse mesh region in sample VG09-12.  We took 50 measurements across the face of VG09-12, including both on- and off- mesh regions.  The measurements are consistent with the predicted transmission, shown as the gray band.  The off-mesh regions demonstrate near-perfect bonding.\label{figVG12sta}}
\end{figure}

We acquired spatially resolved measurements of the pair of bonded parts VG09-12, (Tables \ref{tbl_experiments} and \ref{tbl_meshPatterns}).  We translated the sample through the measurement beam and measured a transmission spectrum at 25 adjacent positions.  We chose a 2 mm step size, which is comparable to the beam width.  We sampled across the mesh pattern and into the off-mesh pattern of sample VG09-12 (Figure \ref{figVG0912_STA_illus}).

The colored dots in the position marked in Figure \ref{figVG0912_STA_illus} match the colors of the spectra in Figure \ref{figVG0912}.  The approximate size of the measurement beam is shown for scale.  The conspicuous mesh pattern has a depth of $49\pm6\;$nm, as measured with an optical profiler, and global fill factor of 50\%.  However, since the beam size is comparable to the mesh grid size, the fill factor for any given measurement can vary from about 30\% to 75\%.  The gray band in Figure \ref{figVG0912} represents the allowable range of predictions, taking into account both the uncertainty in the mesh depth and fill factor.  If the mesh pattern were perfectly bonded, we would expect all the spectra to fall inside the color swath in Figure \ref{figVG0912}.  We see excellent consistency between prediction and measurement, with one measured spectrum hinting at a small gap ($\lesssim15$ nm) in one measurement position.  Outside the mesh area, the bonding is excellent, as demonstrated by the off-mesh transmission with transmission consistent with 100\% to within the measurement uncertainty of 0.2\%.

A key result of these test measurements is to demonstrate that gaps of known dimensions are recovered in our spectroscopy, both for large and small gaps.  Table \ref{tbl_DerivedGapSizes} summarizes the outcome of measurements in the areas with known gaps and fill factors.  The table lists the median sample and 68\% confidence intervals.  The results from the scans shown in Figure \ref{figVG12sta} further support our conclusion.

\subsection{What is the smallest gap this technique can detect?}
\begin{figure}[!htbp]
    \centering
    \begin{subfigure}[b]{0.5\textwidth}
        \includegraphics[width=\textwidth]{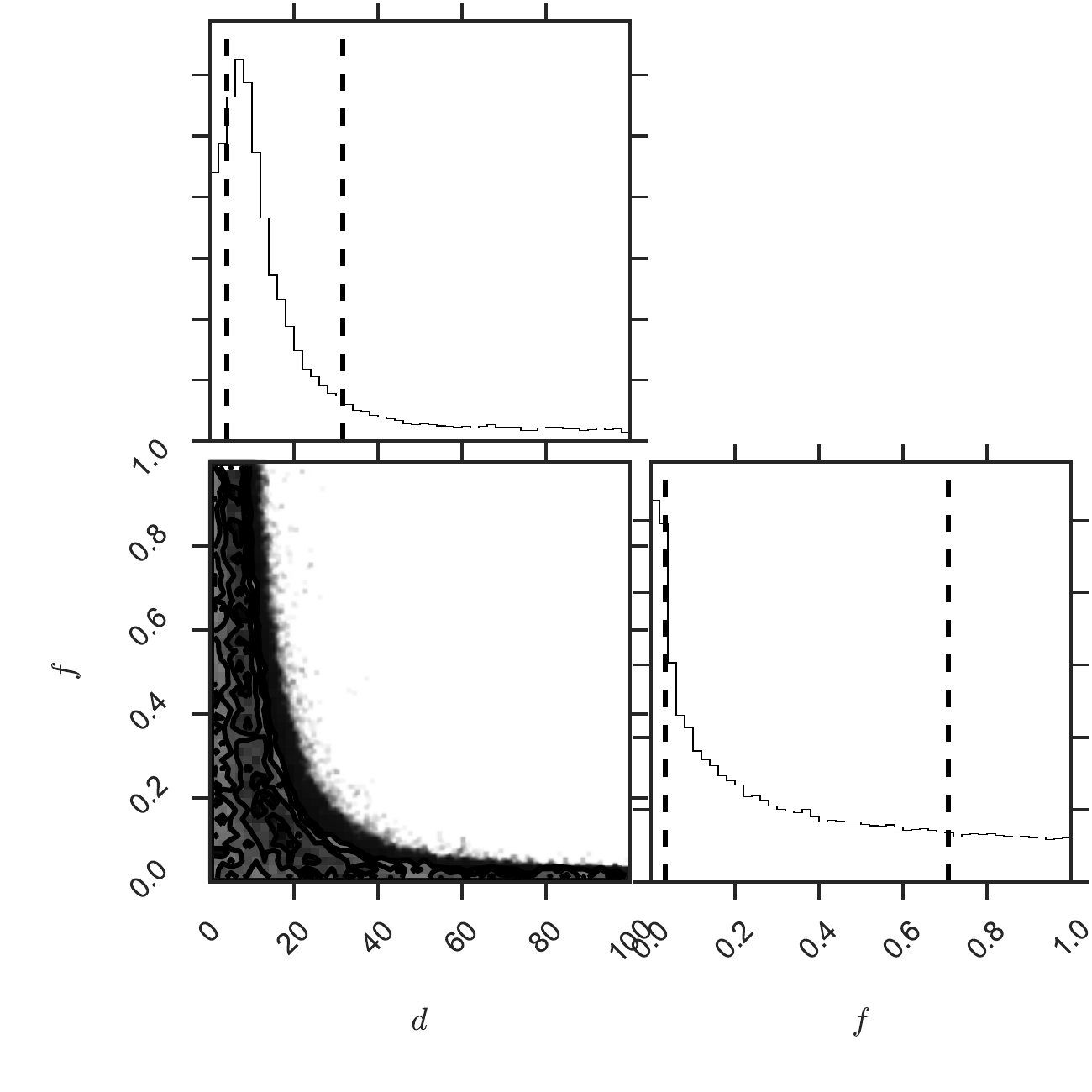}
        \caption{Fitted gap size $d$ and fill factor $f$ for VG09-12 off-mesh}
	\label{VG0912_0gap_corner}
    \end{subfigure}
    \begin{subfigure}[b]{0.5\textwidth}
        \includegraphics[width=\textwidth]{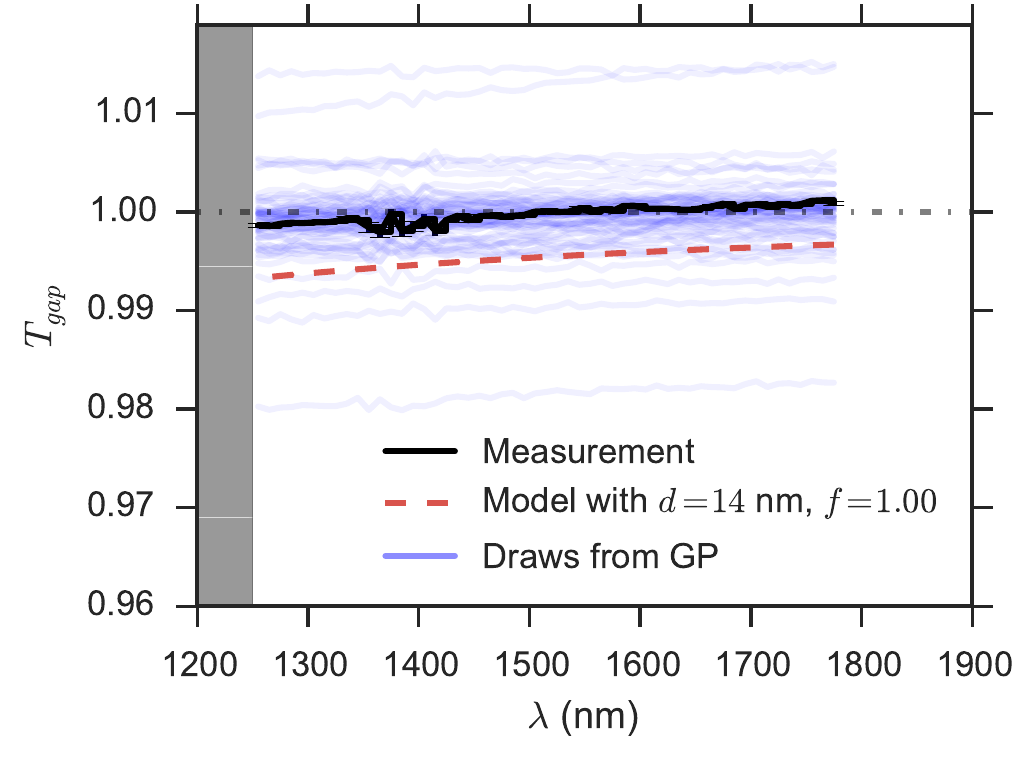}
        \caption{Measured spectrum of VG09-12}
        \label{VG0912_0gap}
    \end{subfigure}
\caption{Similar to Figure \ref{figVG03full} and \ref{figVG03part}, but for sample VG09-12 in the off-mesh region.  The measured spectrum in this region shows no appreciable deficit of transmission from a DSP reference sample.  The blue lines represent individual draws from the Gaussian process model.  
\label{figVG12}}
\end{figure}

We devised two experiments to approach the question, ``What is the smallest gap this technique can detect?''.  We applied our MCMC model fitting formalism to an unbonded DSP wafer.  The measured transmission spectrum was normalized by a transmission spectrum of the same sample taken only a few minutes earlier.  Any difference of this normalized transmission from unity is attributable entirely to measurement uncertainty, and represents the smallest possible signal one could extract.  We also measured bonded sample VG09-12 in a region away from the intentionally implanted gap.  This spectrum was processed in the same way as the spectra of the VG09-12 mesh area, exhibiting a mean value of 0.9998 and a standard deviation $\sigma=0.0008$.  This spectrum is consistent with no gap.  We applied our MCMC formalism to both the unbonded DSP wafer and the VG09-12 off-mesh region.  Table \ref{tbl_DerivedGapSizes} summarizes the formal derivation of gap sizes and fill factors for both the double-side polished puck and the off-mesh region of the VG09-12 bonded pair.  The well-calibrated DSP noise spectrum rules out (at $>3\sigma$) gap sizes larger than about 9.0 nm over $>80$\% of the measurement area.  Similarly, the VG09-12 off-mesh spectrum rules out gaps greater than 14 nm over $>80$\% of the measurement area.  Figure \ref{figVG12} shows the corner plots and spectra for the off-mesh region in VG09-12.

The actual measurement we are making is of the change in transmission through the sample as a function of wavelength.  This change is larger for larger gaps but becomes diluted when the fill factor is less than unity.  This dilution leads to a degeneracy between the axial extent of the gap and its fractional fill factor over the measurement beam,  one that is broken by the harder-to-measure curvature of the transmission spectrum.  Figure \ref{VG0912_0gap_corner} illustrates the degeneracy.  The red dashed line shows the normalized transmission spectrum of a 14 nm gap over 100\% of the measurement area.  Gaps larger than 60 nm would have to cover less than 10\% of the measurement area in order to be consistent with the measurement.

\section{Discussion}

The experiments and measurements in Section \ref{secResults} demonstrate that we can detect and measure the axial extent of gaps down to 14 nm.  The technique is \emph{non-destructive} and can be implemented with any high precision near-IR spectrograph.  The spectrograph need not be high resolution-- our 5 nm spectral resolution was not a limiting factor.  The limiting factor is spectrophotometric \emph{accuracy} and \emph{precision}.  In this Section, we question the assumptions of our technique, identify its limitations, and mention some opportunities.

\emph{More complex underlying gap size distributions-} One key assumption of our gap-as-etalon modeling strategy is that the distribution of gap sizes over the measurement area is bimodal- a gap of size $d$ or zero gap.  This choice was a matter of computational simplicity, and we could easily account for any more complicated gap size distribution by inserting additional terms in Equation \ref{eqnMix}.  For example, if there is a gradient across the bond, the emergent transmission spectrum will be an admixture of many gaps with various areal coverage fractions.  Since the gap sizes for which we intend to use the model are small, and the interface takes place within a subregion of constant gap size, the full solution becomes an areal integral of Equation \ref{eqFP} normalized by the total measurement area.  Solving for the relative distributions of gaps is highly degenerate, in the same way as we saw in the 2-parameter corner plots of gap depth and fill factor.  We experimented with by-eye fitting parametric distributions of gap sizes, characterized by a maximum and minimum gap size, and a monotonic slope of the distribution.  For sub-wavelength gaps, the gap size distribution is too degenerate to derive much useful insight on the slope of the distribution.  Specifically, for $d_{max}<\lambda/4$, our simple mixture model will converge on the spectrum of the average gap size, even if the \emph{True} distribution is more complex.  Reducing the measurement area could break some of these degeneracies if more precision is desired.

\emph{The benefit of the Gaussian process-} We chose to compare models to data with MCMC Gaussian process regression.  Gaussian process regression is not necessary if random errors are much larger than the systematic errors, in which case any model fitting strategy---like least squares or over-plotting a model to data by hand---will suffice.  We quantified the benefit of Gaussian process regression by generating synthetic transmission spectra with known gap size and fill factor, and known covariance structure.  We applied ordinary least squares regression to the synthetic spectra and derived values for $d$ and $f$ that were biased by up to $20\sigma$ or more.  Applying Gaussian process regression recovered the input values to within $1\sigma$.

\emph{Normal incidence- } We have assumed that the transmission measurement is performed at normal incidence.  If the measurement beam forms a non-zero angle with the entrance face, bond interface, or exit face, several things happen to make this technique invalid.  Notably, the polarization effects, Fresnel reflection losses, and the incoherent multiple reflections change the emergent transmission spectrum in non-trivial ways.

\emph{DSP vs AR coated front faces- } We have assumed that the Si$-$air interfaces are approximated by Fresnel reflection losses.  If the entrance and exit faces of the bonded samples are AR coated, the incoherent multiple reflection model will have to be modified.  \emph{Equation \ref{eqn:Tetalon} is not valid if the entrance and exit Si interfaces are AR coated.}  Equation \ref{eqn:SiAirMatrix} would have to be altered with the AR coating's wavelength-dependent transmission, which might not have a simple analytical form.

\emph{How far can this technique go?}  The spectrophotometric precision of our instrument flows down directly to limit the smallest measurable gap size.  This limitation will vary from instrument to instrument.  Our instrument can achieve 10 times better accuracy and precision in ``double beam'' mode than in ``single beam'' mode.  However, the slightly different polarization sensitivity in the different beams caused systematic jumps at the lamp/grating change-over wavelengths.  Agilent offers an adapter to reduce the polarization sensitivity, which mitigates these lamp jumps.  If precisions of 0.02\% could be achieved, gaps $<4$ nm could be reliably measured.  This level begins to be comparable to the effect of interesting physics, like surface roughness and refractive index.  In fact, the van der Waals bond energy could be constrained from these types of measurements \cite{2001JOptA...3...85G}.

\begin{figure}[!htbp]
\includegraphics[width=0.95\columnwidth]{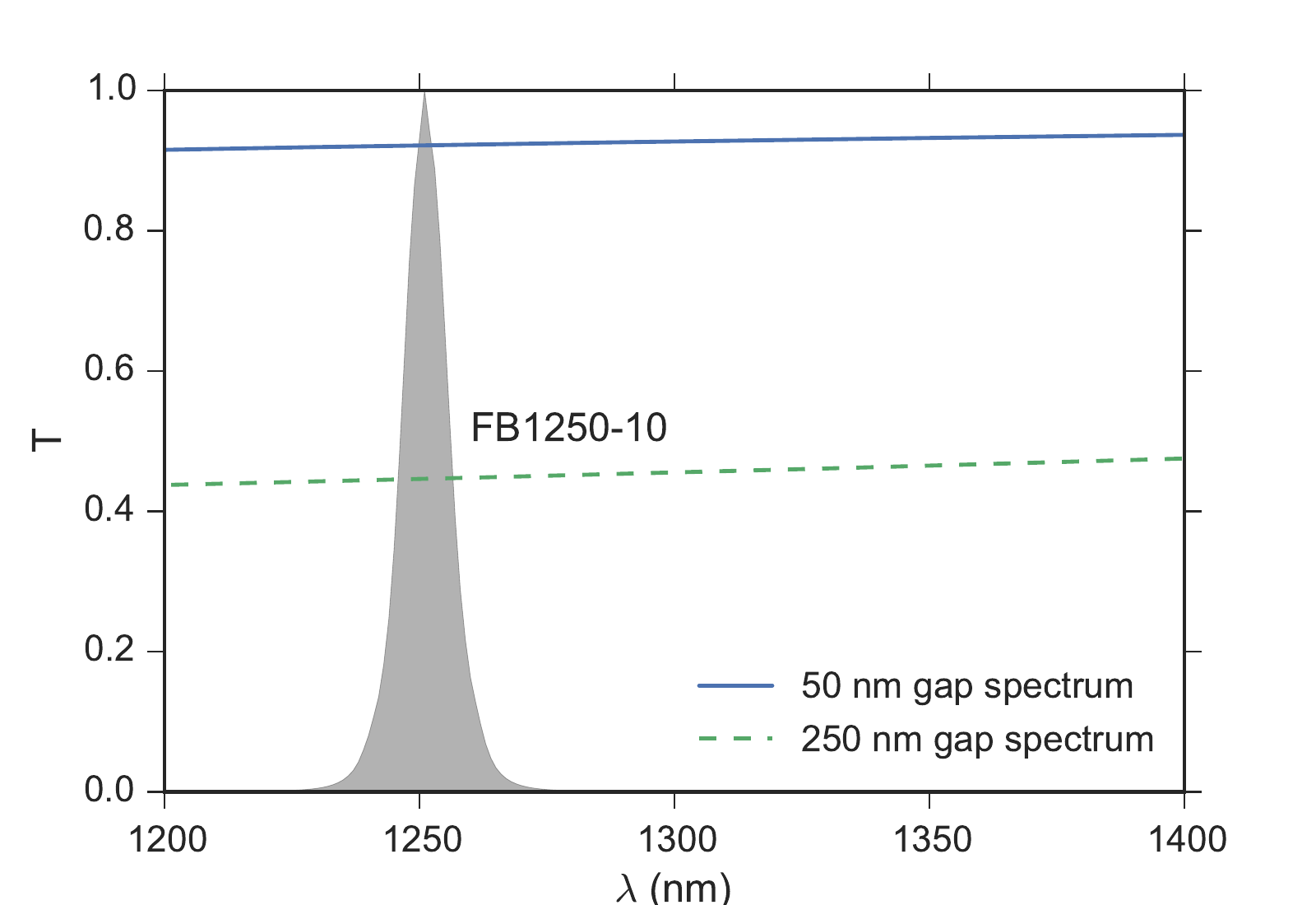}
\caption{\label{figFB1250-10} Filter curve and gap spectra.  The Thorlabs FB1250-10 filter transmission curve is shown as the gray shaded region.  We integrate the gap spectra with this filter curve to relate the deficit of flux in IR imaging with the axial extent of interfacial gaps.}
\end{figure}

\emph{Can this technique be applied to imaging?}  The analysis performed in this article informs smart strategies for using IR imaging for gap detection in bonded Si optics.  In IR images (Figure \ref{figVS2021_IR_image}), interfacial gaps appear as dark patches.  These dark patches are darker at shorter wavelengths as we see from Equation \ref{eqFP}, which shows that observations at short wavelengths have the most sensitivity to small gaps.  Therefore, an IR imaging approach should use the shortest possible wavelength without getting too close to the Si absorption cutoff.  This work recommends a 1250$-$1300 nm filter.  Table \ref{tbl_FB1250} lists the predicted integrated transmission through a commercially available narrow bandpass filter, for example the FB1250-10 from Thorlabs, which has center wavelength 1250 nm and full width at half maximum 10 nm.  We calculated the integrated flux by integrating the product of the etalon transmission (Equation \ref{eqn:Tetalon}) with the FB1250-10 filter curve downloaded from the Thorlabs website\footnote{\url{www.thorlabs.com}}.  The filter curve and example gap spectra are shown in Figure \ref{figFB1250-10}.

\begin{table}[ht]
\caption{FB1250-10 integrated gap transmission \label{tbl_FB1250}}
\begin{center}
    \begin{tabular}{ r r}
    \hline
    $d$ & $T_{1250}$\\
    nm  & - \\
    \hline
      0 &                   1.000 \\
      8 &                   0.998 \\
     16 &                   0.991 \\
     24 &                   0.981 \\
     32 &                   0.966 \\
     40 &                   0.948 \\
     48 &                   0.928 \\
    \hline
    \end{tabular}
\end{center}
\end{table}

\begin{figure}[!htbp]
\includegraphics[width=0.95\columnwidth]{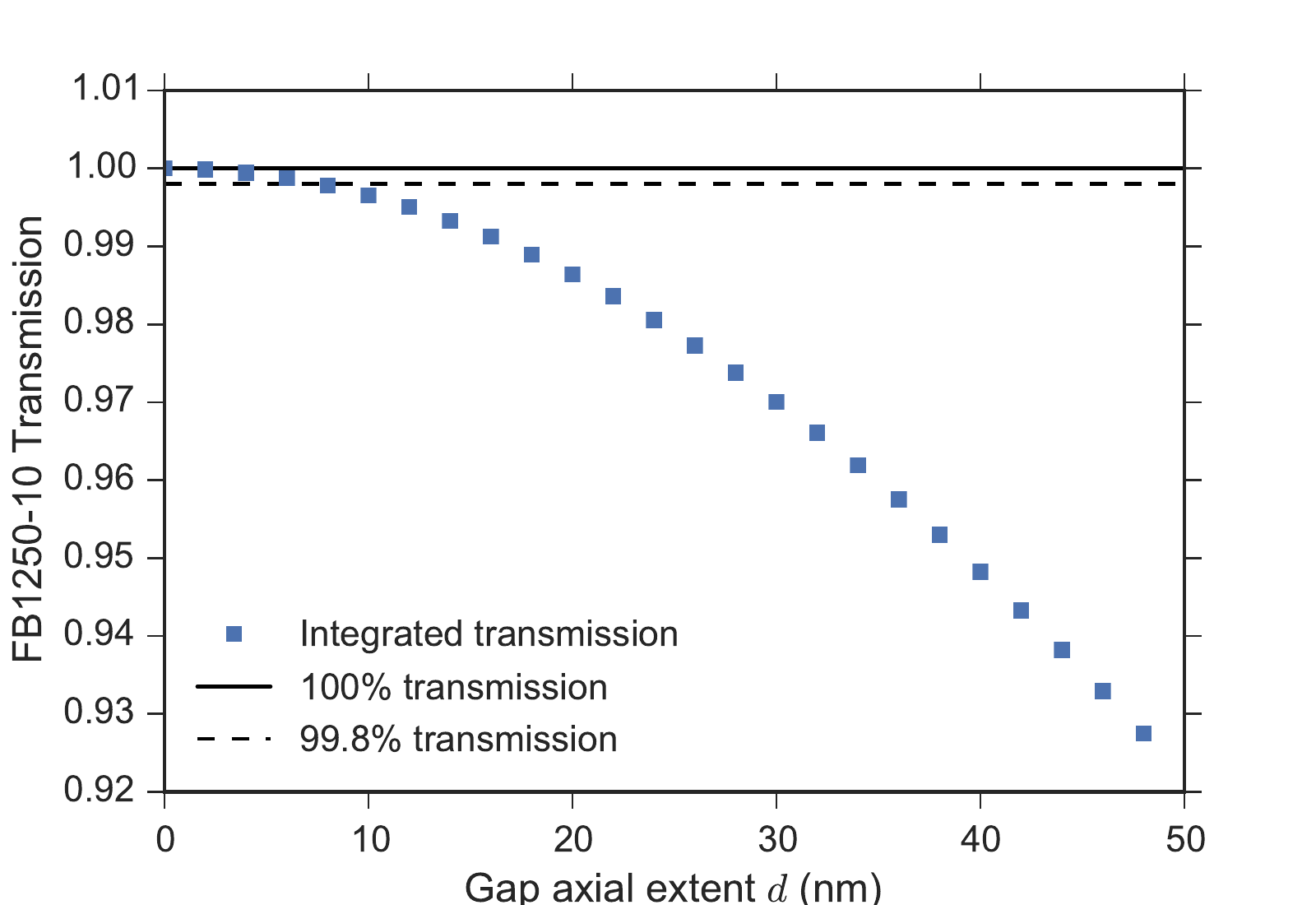}
\caption{\label{figFB1250-10_integ} Integrated transmission through the FB1250-10 filter as a function of interface gap size.  The 99.8\% accuracy achieved in this article is shown as the horizontal dashed line.  Imagers that can achieve 5\% photometric accuracy can detect 40 nm gaps.  This figure assumes uncoated Si substrates.}
\end{figure}

Figure \ref{figFB1250-10_integ} shows the integrated transmission through the FB1250-10 filter for small gaps ($0 < d(\mathrm{nm}) < 50$).  Table \ref{tbl_FB1250} and Figure \ref{figFB1250-10_integ} can be used as a look-up table for converting between the dark patches observed in IR imaging and the axial extent of the interface gap, if the FB1250-10 filter is used during IR imaging.  IR images must be normalized to the transmission through a DSP sample under identical, nearly collimated lighting conditions.  Any deficits in transmission through a bonded sample can be converted to a gap size $d$, using Table \ref{tbl_FB1250}, if the FB1250-10 filter is used.  IR imaging will lack the spectral information available in the spectra shown in this article.  The curvature of the spectra provides assurance of the physical origin of the dark patches seen in IR imaging.  IR imaging will require exceptional quality of flat-fielding or dithering of the images to reach the precision achieved here with relative ease.  

\section{Conclusions}
We have described a measurement technique and laid out an analysis formalism for the detection and measurement of sub-wavelength gaps at Si$-$Si interfaces.  Experimental measurements show that we can precisely recover the extent of wavelength-scale gaps of known size, account for the presence of small gaps, and capture results consistent with no gap when measuring a monolithic part.

The technique and formalism we describe here can detect and measure the axial extent of gaps in Si$-$Si optical bonds down to a level of $\sim14\;$nm.  It makes use of widely available off-the-shelf metrology equipment and simple to implement analysis.  Armed with this technique, optical fabricators have a new diagnostic tool to assess the quality of Si$-$Si bonds.

The authors thank the anonymous referee, whose comments improved the manuscript.  MGS acknowledges support from the NASA Graduate Student Research Program (GSRP) Fellowship, grant NNX10AK82H.  Portions of this research were supported by NASA ACT grant NNX12AC31G.  The authors thank Amanda Turbyfill, Mike Pavel, and Carleen Boyer for assistance with IR imaging and spectroscopy.  MGS thanks the organizers and participants of the \emph{Astro Data Hack Week} (2014), especially Dan Foreman-Mackey for sharing tutorials and code for implementing the Gaussian process technique.  This research has made use of NASA's Astrophysics Data System.  Many of the figures in the paper were made with Matplotlib \cite{Hunter:2007} in IPython Notebooks \cite{PER-GRA:2007}.  The Notebooks with complete version history are available on GitHub\footnote{\url{https://github.com/Echelle/AO_bonding_paper}}.

\appendix

\section{Incoherent Multiple Reflections Transfer Matrix Method}
\label{sec:Append-IMRTMM}

Saleh and Teich (see Chapter 7 ``Fundamentals of Photonics'' \cite{2007fuph.book.....S}) describe the wave transfer matrix method.  Their technique is to assemble a 2$\times$2 scattering matrix $\boldsymbol{S}$ which has the elements:
\begin{eqnarray}
\boldsymbol{S} = \left(
\begin{array}{cc}
 t_{12} & r_{21} \\
 r_{12} & t_{21} \\
\end{array}
\right)
\end{eqnarray}
Where $t$ and $r$ stand for transmission and reflection respectively.  The order of subscripts is the order of origin and destination of the wave with respect to the interface, so we know the origin and direction of the wave.  The $\boldsymbol{S}$ matrix encapsulates all information about how light waves interact with the interface.  The power of the technique comes from the wave transfer matrix, $\boldsymbol{M}$.  The matrix $\boldsymbol{M}$ has the convenient property that its output can be used as the input for another matrix.  In other words, the input vector to $\boldsymbol{M}$ is made of the left and right moving components directly before the interface; the outputs are the left and right moving components directly after the interface:
\begin{eqnarray}
\left(
\begin{array}{c}
 U_2^{(+)} \\
 U_1^{(-)} \\
\end{array}
\right)=\boldsymbol{S} \left(
\begin{array}{c}
 U_1^{(+)} \\
 U_2^{(-)} \\
\end{array}
\right) \\
\left(
\begin{array}{c}
 U_2^{(+)} \\
 U_2^{(-)} \\
\end{array}
\right)=\boldsymbol{M} \left(
\begin{array}{c}
 U_1^{(+)} \\
 U_1^{(-)} \\
\end{array}
\right)
\end{eqnarray}

The elements of $\boldsymbol{S}$ and $\boldsymbol{M}$ are related to each other by geometric transformations\cite{2007fuph.book.....S}.  In thin films, the wavelength is comparable to the size of the dielectric layer and the vector components $U_{i}$ represent the complex amplitudes of the electromagnetic waves.  The polarization state can be encapsulated in scattering matrix components\cite{2007fuph.book.....S}.  The intensities of the emergent spectrum can be computed from the absolute square of the complex amplitudes.  For thick films, the vector components are the intensities of the emergent spectrum, since waves are incoherent.  \cite{2002ApOpt..41.3978K} work out the general transfer-matrix method for optical multilayer systems with incoherent interference.  The key idea for the incoherent transfer matrix method approach is to populate a scattering matrix with elements equal to the (wavelength dependent) transmitted $T_i$ and reflected $R_i$ intensities of an interface or set of interfaces that act together.  Then, use the geometric transformations to construct the $\boldsymbol{M}$ matrix.

The matrix for the silicon air gap is calculated in the following way.  We treat the air gap as a Fabry-P\'{e}rot etalon.  We do not need to consider the microscopic coherent interactions with the etalon transmission, all of that information is encapsulated in these equations for a Fabry-P\'{e}rot etalon model for the gap:

\begin{eqnarray}
 \delta = \frac{2\pi}{\lambda}2d \label{eq:phase} \\
  F \equiv \frac{4R}{(1-R)^2} \\
 T_g = \frac{1}{1+F\sin^2(\delta/2)}  \label{eq:FabPerot}
\end{eqnarray}

with $\lambda$ the vacuum wavelength, $R$ the Fresnel reflection of silicon, and $F$ the coefficient of finesse.  The coefficient of finesse $F$ and the phase $\delta$ are the two parameters of the Fabry-P\'{e}rot etalon.  The coefficient of Finesse encapsulates the Fresnel reflection and depends only on the Si refractive index which has only a small wavelength (and temperature) dependence.  The phase depends (Equation \ref{eq:phase}) on the wavelength $\lambda$ and $d$ the air gap spacing.  We assume the gap is lossless, i.e. $T_g+R_g=1$.  The incoherent scattering and transfer matrices for the gap are then:

\begin{eqnarray}
\boldsymbol{S_g}&=&\frac{1}{1+F\sin^2{\delta/2}} \left(
\begin{array}{cc}
1 & F \sin ^2(\delta/2) \\
F \sin ^2(\delta/2) & 1 \\
\end{array}
\right) \nonumber \\
\nonumber \\
\boldsymbol{M_g}&=&\left(
\begin{array}{cc}
 1-F \sin ^2(\delta/2) & F \sin ^2(\delta/2) \\
 -F \sin ^2(\delta/2) & 1+F \sin ^2(\delta/2) \\
\end{array}
\right)
\label{eqn:EtalonMatrix}
\end{eqnarray}

We assembled the matrix for the Air-Si Fresnel interface at the exterior of the bonded Si parts in the following way.  First it is important to note that the matrix is the same whether the transmission is from Si to air or air to Si.  This reciprocity is not necessarily true for the complex amplitudes matrix, but our approach employs intensities not complex amplitudes.  The Fresnel interface is lossless.  The transmission and reflection are given by the Fresnel equation for normal incidence:
\begin{eqnarray}
T_n&=&\frac{4n_{Si}}{(n_{Si}+1)^2} \\
R_n&=&\frac{(n_{Si}-1)^2}{(n_{Si}+1)^2} \label{eq:FresnelTrans}
\end{eqnarray}
For clarity we will drop the subscripts from $n_{Si}$, since we have already set $n_{air}=1$ and there are no other dielectric interfaces to think about.  The scattering and transfer matrices for the air-Si Fresnel boundary are then:

\begin{eqnarray}
\boldsymbol{S_n}&=&\frac{1}{(n+1)^2} \left(
\begin{array}{cc}
4n & (n-1)^2 \\
(n-1)^2 & 4n \\
\end{array}
\right)  \nonumber \\
\nonumber \\
\boldsymbol{M_n}&=&\frac{1}{4n}\left(
\begin{array}{cc}
 -n^2+6  n-1 & ( n-1)^2 \\
 -( n-1)^2 & ( n+1)^2 \\
\end{array}
\right)
\label{eqn:SiAirMatrix}
\end{eqnarray}

Finally, we cascade the matrices together to compute the net transmission through the stack of abstractions.  The result is a $2\times2$ transfer matrix:

\begin{eqnarray}
\boldsymbol{M_{net}}=\boldsymbol{M_n}\boldsymbol{M_g}\boldsymbol{M_n}
\end{eqnarray}

From the matrix transformation equations \cite{2007fuph.book.....S} we know $M_{22}=1/T$.  Taking the inverse of the bottom right element of $\boldsymbol{M_{net}}$, we get the transmission $T_{net}$ through the net optical device:

\begin{eqnarray}
T_{net}=\frac{2 n}{1+ 2n F\sin ^2(\delta/2)+n^2} \label{eqn:FPmatTrans}
\end{eqnarray}

The matrix technique recovers the usual relation for transmission through a double-side polished dielectric (Equation \ref{eqnAbsDSPtrans}).  For this case, the matrix multiplication is simply $\boldsymbol{M_{DSP}}=\boldsymbol{M_n}\boldsymbol{M_n}$.  Taking the inverse of the $M_{22}$ element, we find:

\begin{eqnarray}
T_{DSP}=\frac{2 n}{n^2+1}\label{eqn:EqofSummedSlab}
\end{eqnarray}

which is identical to the result obtained by directly summing the intensities from multiple reflections:
\begin{eqnarray}
T_{DSP}=T^2 \sum_{i=0}^{N}R^{2i} \label{eqn:multsum}
\end{eqnarray}

It is informative to isolate the effect of the gap by dividing the measured bonded wafer transmission by the transmission of a reference DSP Si part.  We call this normalized transmission the gap transmission $T_{g}$:
\begin{eqnarray}
T_{g} = T_{net}/T_{DSP} \\
T_{g} = \frac{n^2+1}{2 n F \sin ^2(\delta/2)+n^2+1} \label{eqn:Tetalon}
\end{eqnarray}

\bibliographystyle{osajnl}

\end{document}